# THERMAL FLUCTUATIONS IN THE LANDAU-LIFSHITZ-BLOCH MODEL

*Marco Menarini*[*1] *and Vitaliy Lomakin*[1]

[1] Department of Electrical and Computer Engineering, Center for Memory and Recording Research, University of California, San Diego, La Jolla, California 92093-0319

(Dated: April 17, 2020)

A formulation for thermal noise in the stochastic form of the Landau-Lifshitz-Bloch equation used for modeling the magnetization dynamics at elevated temperatures is presented. The diffusion coefficients for thermal fluctuations are obtained via the Fokker-Plank equation using the mean field approximation of the field in defining the free energy. The presented model leads to a mean magnetization consistent with the equilibrium magnetization for small and large particles. The distribution of the magnetization magnitude is of the Maxwell-Boltzmann type. The presented model was tested by studying the equilibrium magnetization in macrospin particles at high temperatures. The model is appealing for multi-scale modeling, such as modeling heat assisted magnetic recording systems and all-optical magnetization reversal.

## I. INTRODUCTION

Understanding the magnetization dynamics at high temperature is important for our fundamental understanding of nanomagnetism and for a set of applications, such as heat assisted magnetic recording technologies (HAMR) [1-3] and ultrafast optical processes [4-7]. Numerically modeling such systems is complicated because the material properties change significantly at elevated temperatures, especially at temperatures near or above the Curie temperature.

Atomistic spin models have been used to provide parameterization of thermal properties, such as the equilibrium magnetization $m_e(T)$, anisotropy, and susceptibility [8]. The atomistic models, however, are not fit to simulate large-scale systems, such as those of common interest in magnetic recording and opto-magnetic simulations [8]. To solve this problem, several micromagnetic models have been proposed that use a macrospin to represent the behavior of an ensemble of atoms. The main idea behind the macrospin model is to use a single vector to represent the assembly of a large number of atoms in a finite volume. The field acting on this magnetization vector is obtained from the atomistic Hamiltonian via the mean field approximation [9,10]. In micromagnetic models, the average magnetization of the system at a certain temperature is described by the equilibrium magnetization obtained from the atomistic model. Additionally, elevated temperatures result in thermal noise. The introduction of stochastic fluctuations that correctly model the behavior of this noise is of practical importance to study the magnetization behavior, such as reversal time, signal to noise ratio, and jitter in HAMR [10].

Several stochastic formulations of the Landau-Lifshitz-Bloch (LLB) equation have been proposed to introduce the thermal fluctuations in the model [11-13]. However, these models have limitations, e.g. they may ignore the fluctuations on the longitudinal component of the magnetization [13], underestimating the longitudinal fluctuations contribution, or lead to a mean magnetization higher than the expected equilibrium magnetization [11,12]. These limitations may make such models hard to use for multiscale modelling.

---

[*] menarini.marco@gmail.com

In this paper, we introduce an alternative form of the thermal noise in the stochastic LLB equation, which is consistent with the solution of the Fokker-Planck (FP) equation. This form preserves the magnitude of the equilibrium magnetization without ignoring the stochastic contribution on the longitudinal component. The model is based on the formalism introduced by Garcia-Palacios [14] and Garanin [11] but it uses a different free energy definition for introducing the thermal fluctuations. The model is validated against other existing formulation of the stochastic LLB model by considering the distribution of the magnetization at equilibrium for macrospin particles.

## II. THE STOCHASTIC LANDAU-LIFSHITZ-BLOCH EQUATION

Our aim is to construct a mathematical consistent model for thermal fluctuations in the LLB equation applicable for numerical simulations. We start by introducing the LLB formulation and by defining the free energy of the system. We then introduce the Langevin equation to augment the deterministic LLB equation with stochastic components. Finally, we obtain the strength of diffusion coefficients of the thermal noise by solving the Fokker-Plank equation for the Langevin form of the LLB.

### A. THE LLB EQUATION FOR FERROMAGNETS

The starting point of our derivation is the original LLB equation for ferromagnets by Garanin under the assumption of small deviation from the equilibrium (see Eq. (2.17) in Ref [9]). This LLB formulation is based on the classical Hamiltonian assuming biaxial anisotropic exchange interactions and an applied field

$$\mathcal{H} = -\mu_0 \sum_i \mathbf{H}_i \mathbf{s}_i - \frac{1}{2} \sum_{ij} J_{ij} \left( \eta_x s_{xi} s_{xj} + \eta_y s_{yi} s_{yj} + s_{zi} s_{zj} \right), \qquad (1)$$

where $\mu_0$ is the magnetic moment of the atom, $\mathbf{H}_i$ is the external magnetic field acting on the spin $\mathbf{s}_i$, $J_{ij}$ is the exchange integral between atom $i$ and $j$, and $\eta_x \leq \eta_y \leq 1$ are the $x$ and $y$ anisotropy coefficients. Under the assumption of a small-anisotropy field compared to the exchange integral, which is valid for most ferromagnets, i.e. $\eta'_{x,y} = 1 - \eta_{x,y} \ll 1$, it is possible to express the magnetic field acting on a macrospin using the mean field approximation in the form:

$$\begin{aligned}
\mathbf{H}^{MFA} &= \mathbf{H} + \frac{J_0}{\mu_0} \left[ \alpha \Delta \mathbf{m} - \eta'_x \mathbf{m}_x - \eta'_y \mathbf{m}_y \right] + \frac{J_0}{\mu_0} \mathbf{m} = \mathbf{H}'_{eff} + \mathbf{H}_E, \\
\mathbf{H}'_{eff} &= \mathbf{H} + A_{exc} \Delta \mathbf{m} - \frac{J_0}{\mu_0} \left( \eta'_x \mathbf{m}_x + \eta'_y \mathbf{m}_y \right), \\
\mathbf{H}_E &= \frac{J_0}{\mu_0} \mathbf{m},
\end{aligned} \qquad (2)$$

where $J_0$ is the zero-order Fourier component of the exchange interaction and $A_{exc} = \alpha J_o / \mu_0$ is the micromagnetic exchange coefficient with $\alpha = a^2 / z$, where $a$ is a lattice dependent constant, e.g. with $a = 3.8 \text{ Å}$ for FePt, and $z$ is the average number of nearest neighbor in the lattice.

Let us define an instantaneous equilibrium magnetization, as the thermal equilibrium value for a given reduced magnetic field $\xi_0$ [9]:

$$\mathbf{m}_0 = B_s(\xi_0)\boldsymbol{\xi}_0 / \xi_0, \tag{3}$$

where $\boldsymbol{\xi}_0 = \mu_0 \mathbf{H}^{MFA}/T$ is the reduced magnetization and $B_s$ is the Brillouin function. Assuming small deviations of the magnetization from this equilibrium state, one can write the LLB equation for ferromagnets in the following form

$$\frac{d\mathbf{m}}{dt} = \gamma \left[\mathbf{m} \times \mathbf{H}'_{eff}\right] - \Gamma_1 \left(1 - \frac{\mathbf{m} \cdot \mathbf{m}_0}{m^2}\right)\mathbf{m} - \Gamma_2 \frac{\left[\mathbf{m} \times \left[\mathbf{m} \times \mathbf{m}_0\right]\right]}{m^2}, \tag{4}$$

where $\gamma$ is the gyromagnetic ratio, $B'_s$ is the derivative of the Brillouin function with respect to its argument, $\Gamma_1 = \Lambda_N B_s(\xi_0)/(\xi_0 B'_s(\xi_0))$ and $\Gamma_2 = \Lambda_N \left(\xi_0/B_s(\xi_0) - 1\right)/2$ are the longitudinal and transverse relaxation rates, respectively, $\Lambda_N = 2\gamma\lambda T/\mu_0$ is the Neel attempt frequency with $\lambda \leq 1$, the atomistic damping coefficient. Assuming that $\mathbf{H}'_{eff}$ is in the direction of $\mathbf{m}_0$, we can re-write Eq. (4) as

$$\frac{d\mathbf{m}}{dt} = \gamma \left[\mathbf{m} \times \mathbf{H}'_{eff}\right] - \gamma\tilde{\alpha}_{\parallel}\left(\frac{1 - B_s(\mu_0\beta H^{MFA})/m}{\mu_0\beta B'_s(\mu_0\beta H^{MFA})}\right)\mathbf{m} - \gamma\tilde{\alpha}_{\perp}\frac{\left[\mathbf{m} \times \left[\mathbf{m} \times \mathbf{H}'_{eff}\right]\right]}{m^2}, \tag{5}$$

where $\beta = 1/T$ and $\tilde{\alpha}_{\parallel} = 2\lambda T/\tilde{J}_0$ and $\tilde{\alpha}_{\perp} = \lambda(1 - T/\tilde{J}_0)$ are the unitless longitudinal and transverse damping parameters, respectively, with a modified coefficient $\tilde{J}_0 = J_0 + \mu_0 \mathbf{m} \cdot \mathbf{H}'_{eff}/m^2$.

Assuming that $|H_E| \gg |H'_{eff}|$, which is typically realized as long at the temperature is not very close to $T_C$, and Taylor expanding the instantaneous equilibrium magnetization $\mathbf{m}_0$ up to the first order around $\mathbf{H}'_{eff}$ the LLB equation assumes the simplified form:

$$\frac{d\mathbf{m}}{dt} = \gamma \left[\mathbf{m} \times \mathbf{H}'_{eff}\right] - \gamma\alpha_{\parallel}\left(\frac{1 - B_s(m\beta J_0)/m}{\mu_0\beta B'_s(m\beta J_0)} - \frac{\mathbf{m} \cdot \mathbf{H}'_{eff}}{m^2}\right)\mathbf{m} - \gamma\alpha_{\perp}\frac{\left[\mathbf{m} \times \left[\mathbf{m} \times \mathbf{H}'_{eff}\right]\right]}{m^2}, \tag{6}$$

and $\alpha_{\parallel} = 2\lambda T/J_0$ and $\alpha_{\perp} = \lambda(1 - T/J_0)$. The LLB equation (6) is the form that is typically used form micromagnetic modeling [9,11,12,15]. In the following derivations we, however, use the form of Eq. (5) since it is more convenient and general.

For the following derivations we need to define the free energy of the system. To this end we note that in Eqs. (4)-(6), the field used in the precessional term is $\mathbf{H}'_{eff}$ instead of $\mathbf{H}^{MFA}$ because, by construction, $\mathbf{m} \times \mathbf{H}_E = 0$. The magnetic field $\mathbf{H}^{MFA}$ does not appear explicitly in the numerator of the longitudinal relaxation term, but, instead, it is used as an input to the Brillouin function. Furthermore, the first element between parenthesis in the direction of the magnetization in Eq. (6) is not a part of $H^{MFA}$. As a result, the free energy of the system is defined as

$$F(\mathbf{m}) = VM_S^0 \mathbf{H}^{MFA} \cdot \mathbf{m}, \tag{7}$$

where $V$ is the macrospin volume and $M_S^0$ is the saturation magnetization at zero temperature. This definition of the free energy is consistent with the original definition by Brown considering $\mathbf{H}^{MFA}$ as the molecular field. It is also consistent with the definition of the free energy used by Xu and Zhang [13] and by Tzoufras and Grobis [16] in their model for the magnetization dynamics at elevated temperatures.

With respect to the instantaneous equilibrium magnetization $\mathbf{m}_0$, we note that it is a function of the temperature and field. We also can define the equilibrium magnetization $m_e$, which is obtained via the equation [9]:

$$m_e(T) = B_s(\beta J_0 m_e) \ . \tag{8}$$

This equilibrium magnetization is different from the instantaneous equilibrium magnetization defined in Eq. (3) in that it is not a function of the field and it depends only on the temperature via the molecular field approximation $H_E$. For an isotropic particle in the absence of an effective field, the equilibrium magnetization $m_e$ is identical to $m_0$ but it may be different when an effective field is present. The definition of $m_e$ is important when defining the material parameters as it is closely related to the temperature dependent saturation magnetization, e.g. as defined in conventional approached in the LLG equation.

## B. THE STOCHASTIC DIFFERENTIAL EQUATIONS

Starting from Eq. (5), we construct the Langevin form by introducing the stochastic fluctuations in the three perpendicular components of Eq. (5) as an additive term to the field, including precession, longitudinal relaxation, and transverse relaxation components. This leads to three different multiplicative components in the equation due to the cross and dot products components [14]. We can, then, write the Langevin form of the magnetization dynamics equation as

$$\frac{dm_i}{dt} = A_i(\mathbf{m},t) + \sum_k B_{ik}^{(0)}(\mathbf{m},t) L_k^{(0)}(t) + \sum_k B_{ik}^{(1)}(\mathbf{m},t) L_k^{(1)}(t) + \sum_k B_{ik}^{(2)}(\mathbf{m},t) L_k^{(2)}(t) , \tag{9}$$

$$A_i = \gamma \left[\mathbf{m} \times \mathbf{H}'_{eff}\right]_i - \gamma \tilde{\alpha}_\| \left(\frac{1 - B_s(m\beta \tilde{J}_0)/m}{\mu_0 \beta B'_s(m\beta \tilde{J}_0)}\right) m_i - \gamma \tilde{\alpha}_\perp \frac{\left[\mathbf{m} \times \left[\mathbf{m} \times \mathbf{H}'_{eff}\right]\right]_i}{m^2} , \tag{10}$$

$$B_{ik}^{(0)}(\mathbf{m},t) = \gamma \sum_j \epsilon_{ijk} m_j \qquad B_{ik}^{(1)}(\mathbf{m},t) = \frac{\gamma \tilde{\alpha}_\|}{m^2} m_i m_k \qquad B_{ik}^{(2)}(\mathbf{m},t) = \frac{\gamma \tilde{\alpha}_\perp}{m^2}\left(m^2 \delta_{ik} - m_i m_k\right), \tag{11}$$

where $\mathbf{m} = (m_x, m_y, m_z)$ with $i = x,y,z$, $\epsilon_{ijk}$ is the Levi-Civita symbol defining the totally antisymmetric unit tensor, and $\delta_{ij}$ is the delta function. The "Langevin" sources are modelled as Wiener stochastic processes and are assumed to be (i) Gaussian with zero mean, (ii) stationary, (iii) and such that $L_i(t)$ and $L_j(t+\tau)$ are correlated only for a time interval $\tau$ that is much shorter than the time required to observe an appreciable change in the magnetization, i.e. we assume that the collision time between spins is much shorter than the micromagnetic relaxation time [14,17,18]. Under these assumptions the Langevin sources can be written as

$$\left\langle L_k^{(v)}(t) \right\rangle = 0, \qquad \left\langle L_k^{(v)}(t), L_l^{(v)}(s) \right\rangle = 2 D_v \delta_{kl} \delta(t-s) , \tag{12}$$

where $D_v$ with $v = 0,1,2$ are the diffusion coefficient to be determined by solving the corresponding FP equation at equilibrium.

## C. FOKKER-PLANCK EQUATION

The time evolution of the transitional probability density function $f(m,t|m_0,t_0)$ governing the magnetization can be obtained by solving the Fokker-Planck equation associated to the Langevin equation (9). Since the noise enters in the system in a multiplicative way, the correct Langevin equation can been solved using the Stratonovich calculus to obtain the correct thermal equilibrium properties [14]. Using the Stratonovich calculus it is possible to write the FP equation in the form of a continuity equation for the probability density $f$:

$$\frac{\partial f}{\partial t} = -\sum_i \frac{\partial}{\partial m_i} \left\{ \left[ A_i - \sum_{v=0}^{2} D_v \left( \sum_k B_{ik}^{(v)} \left( \sum_j \frac{\partial B_{jk}^{(v)}}{\partial m_j} \right) - \sum_k B_{ik}^{(v)} B_{jk}^{(v)} \frac{\partial}{\partial m_j} \right) \right] f \right\}. \tag{13}$$

Using Eq. (9) in Eq. (13), and noticing that $\sum_k B^{(v)}_{ik} \sum_j (\partial B^{(v)}_{jk}/\partial m_j) = 0$ for $v = 0, 2$ and $\sum_k B^{(1)}_{ik} \sum_j (\partial B^{(1)}_{jk}/\partial m_j) = 2 D_2 \alpha_\parallel^2 \gamma^2 m_i f / m^2$, it is possible to rewrite the Fokker-Planck equation in a more explicit form:

$$\frac{\partial f}{\partial t} =$$
$$-\frac{\partial}{\partial \mathbf{m}} \cdot \left\{ \gamma \left[ \mathbf{m} \times \mathbf{H}'_{eff} \right] f - \gamma \tilde{\alpha}_\parallel \left( \frac{1 - B_S(m\beta \tilde{J}_0)/m}{\mu_0 \beta B'_S(m\beta \tilde{J}_0)} \right) \mathbf{m} f - \gamma \tilde{\alpha}_\perp \frac{\left[ \mathbf{m} \times \left[ \mathbf{m} \times \mathbf{H}'_{eff} \right] \right]}{m^2} f \right\}$$
$$-\frac{\partial}{\partial \mathbf{m}} \cdot \left\{ \gamma^2 \left( \frac{\tilde{\alpha}_\perp^2 D_2}{m^2} + D_0 \right) \left[ \mathbf{m} \times \left[ \mathbf{m} \times \frac{\partial f}{\partial \mathbf{m}} \right] \right] - \gamma^2 \frac{\tilde{\alpha}_\parallel^2 D_1}{m^2} \mathbf{m} \left( \mathbf{m} \cdot \frac{\partial f}{\partial \mathbf{m}} \right) \right\}$$
$$-\frac{\partial}{\partial \mathbf{m}} \cdot \left\{ 2 D_1 \tilde{\alpha}_\parallel^2 \gamma^2 \frac{\mathbf{m} f}{m^2} \right\}. \tag{14}$$

This form is similar to the one derived by Evans et al. [12] with the main difference being the definition of the effective field in the longitudinal component as discussed in Sec. II.A.

Equation (14) has several important properties. Solving it under the stationary condition, i.e. for $\partial f/\partial t = 0$, we can obtain the diffusion coefficients. Due to the presence of the drift (last) term in Eq. (14), one can conclude that the FP should have a solution in the form of the Maxwell-Boltzmann (MB) like distribution

$$f(\mathbf{m}) = f_0 m^2 \exp\left( -\frac{F(\mathbf{m})}{k_B T} \right) = f_0 m^2 \exp\left( -\frac{M_S^0 V \left( \mathbf{H}^{MFA} \cdot \mathbf{m} \right)}{k_B T} \right), \tag{15}$$

where $F(\mathbf{m})$ is the free energy defined in Eq. (7), $k_B$ is the Boltzmann constant, and $f_0$ is a scaling factor. The MB like distribution is common to describe the distribution of a vector length in the presence of an external fluctuation source, e.g. the wind speed in many wind power generation models [19]. It is also consistent with the distribution of the magnetization observed in magnetic resonance imaging experiments [20,21].

To further understand this behavior, we can consider the magnetization length as the sum of a discrete population of $N$ spins that assume the states $S = \pm 1$

$$m = \frac{1}{N}\left|\sum_{i=1}^{N} S_i\right| . \tag{16}$$

When the temperature is low, such as $T \ll T_C$, and due to the strong exchange coefficient $J_0$, most of the spins are aligned in the same direction. The thermal fluctuations randomly flip the spins and the resulting magnetization length distribution appears as the Boltzmann distribution with a narrow standard deviation. When the temperature is above the Curie temperature (i.e. $T > T_C$), the spin up and down populations are almost the same, providing a value of the equilibrium magnetization close to zero and a wider standard deviation. Since the magnetization length is a positive number, the spin flipping cannot produce a negative magnetization. Moreover, the probability of producing a magnetization magnitude below the equilibrium value is lower than that of producing a greater magnetization, which is consistent with Eq. (16). The deviation from the Boltzmann distribution is stronger for smaller particles (practically particles of size smaller than 10 nm). For larger particles or at low temperatures, this deviation would be much less significant, and the contributions of the longitudinal fluctuations become negligible. It is important to note, however, that in various applications, e.g. HAMR, the dimension of the grains is of the order of $5-8$ nm [3,22], and a correct assessment of the noise in this range is important for providing quantitative and qualitative information on the contribution of the noise. Moreover, the intensity of the longitudinal noise can influence the intensity of the optical source necessary to describe the optical reversal in ferromagnetic and ferrimagnetic media [7,23,24].

We, then, note that the contributions of the precessional thermal fluctuation $D_0$ and the transverse relaxation $D_2$ act on the same direction suggesting a correlation between the two fluctuators. To avoid any correlation, we set $D_0 = 0$. This choice is arbitrary and a different choice for the diffusion coefficient can lead to equivalent stochastic processes. This point is shown in Ref. [14] for different implementations of the thermal fluctuations in the Landau-Lifshitz (LL) model, i.e. thermal fluctuations only in the precessional term or in both precessional and transverse term. The diffusion coefficient in such equivalent choices differs only by a scaling factor.

Since the Langevin noise acting on the longitudinal component of the magnetization $D_1$ and on the transverse component of the magnetization $D_2$ are perpendicular to each other by construction of the FP equation, the condition of uncorrelated fluctuations for $(\mathbf{m} \cdot \mathbf{L}^{(1)})\mathbf{m}$ and $\mathbf{m} \times (\mathbf{m} \times \mathbf{L}^{(2)})$ is automatically satisfied. The diffusion coefficient can, then, be obtained by solving the uncoupled system of equation

$$\gamma^2 \frac{\tilde{\alpha}_\parallel^2 D_1}{m^2} \mathbf{m}\left(M_S^0 V \frac{\mathbf{m} \cdot \mathbf{H}^{MFA}}{k_B T} - 2\frac{\mathbf{m} \cdot \mathbf{m}}{m^2}\right) f = \gamma \tilde{\alpha}_\parallel f \left(\frac{1 - B_S(m\beta\tilde{J}_0)/m}{\mu_0 \beta B_S'(m\beta\tilde{J}_0)}\right) \mathbf{m} - 2D_1 \tilde{\alpha}_\parallel^2 \gamma^2 \frac{\mathbf{m}}{m^2} f , \tag{17}$$

$$\gamma^2 \frac{\tilde{\alpha}_\perp^2 D_2}{m^2} \mathbf{m} \times \left[\mathbf{m} \times \left(M_S^0 V \frac{\mathbf{H}'_{eff} + \mathbf{H}_E}{k_B T} - 2\frac{\mathbf{m}}{m^2}\right)\right] f = \gamma \tilde{\alpha}_\perp \frac{[\mathbf{m} \times [\mathbf{m} \times \mathbf{H}'_{eff}]]}{m^2} f . \tag{18}$$

With the MB distribution, the last term in the left- and right-hand sides of Eq. (17) are canceled out and the longitudinal component diffusion coefficient becomes

$$D_1 = \frac{k_B T}{\gamma \tilde{\alpha}_\parallel M_S^0 V} \frac{\mu_0}{\tilde{J}_0} \left(\frac{1 - B_S(m_0 \beta \tilde{J}_0)/m_0}{\mu_0 \beta B_S'(m_0 \beta \tilde{J}_0)}\right) . \tag{19}$$

Defining a scaling coefficient as

$$\eta = \frac{\mu_0}{\tilde{J}_0}\left(\frac{1 - B_s(m_0\beta\tilde{J}_0)/m_0}{\mu_0\beta B'_s(m_0\beta\tilde{J}_0)}\right), \qquad (20)$$

we can write the diffusion coefficient for the longitudinal component as:

$$D_1 = \frac{k_B T}{\gamma \tilde{\alpha}_\| M_s^0 V}\eta, \qquad (21)$$

Using the definition of $m_0$ in Eq. (3), we can show that for $T < T_c$ the effect of the thermal fluctuation due to $D_1$ is negligible, i.e. $\eta$ can be set to zero. For $T > T_c$, the contribution of the thermal fluctuation due to $D_1$ is not negligible. To obtain this contribution, we keep the dominant terms in the Taylor expansion of $B_s$ and $B'_s$ around 0 in Eq. (20), leading to

$$\eta = \frac{1 - \beta\tilde{J}_0 C(S)}{\beta\tilde{J}_0 C(S)} \approx \frac{T}{T_c} - 1, \qquad (22)$$

where $C(S) = (S+1)/(3S)$ and the latter approximation is obtained by assuming that $\tilde{J}_0 \approx J_0 = 3k_b T_c/(S(S+1))$, $\beta = S^2/(k_b T)$. Using Eq. (22) and (19), and assuming that $\tilde{\alpha}_\| \approx \alpha_\|$ for $|H_E| \gg |H'_{eff}|$, we can rewrite the diffusion coefficient for the longitudinal component as:

$$D_1 \approx \begin{cases} 0 & \text{for } T < T_c \\ \dfrac{k_B T}{\gamma \alpha_\| M_s^0 V}\left(\dfrac{T}{T_c} - 1\right) & \text{for } T \geq T_c \end{cases}. \qquad (23)$$

We note that the result given in Eq. (23) satisfies the Fluctuation-Dissipation theorem at low temperatures (see Appendix).

In Eq. (18), we note that the field **H**$_E$ is parallel to the direction of the magnetization **m** by construction and its contribution to the cross-product vanishes. Thus, the diffusion coefficient for the transverse fluctuations can be written as

$$D_2 = \frac{k_B T}{\gamma \tilde{\alpha}_\perp M_s^0 V}, \qquad (24)$$

where $\tilde{\alpha}_\|$ can also be approximated as $\tilde{\alpha}_\| \approx \alpha_\|$ for $|H_E| \gg |H'_{eff}|$. The diffusion coefficient $D_2$ is the same as obtained in previous works [11], and it is equivalent to the diffusion coefficient obtained for the Landau-Lifshitz-Gilbert equation [14].

The formulation for the fluctuations in Eqs. (23) and (24) has similarities and differences as compared to the other formulations [11,12]. Compared to the original LLB formulation [11], the formulations differ in the value of $\eta = 1$, which is greater in the original LLB formulation for the longitudinal fluctuations. The greater longitudinal fluctuations may lead to underestimation of the mean value of the magnetization as compared to the equilibrium magnetization $m_e$. The stochastic LLB formulation by Evans [12] introduces not only a multiplicative but also additive noise [12]. Due to the additive noise acting in all direction this formulation requires introducing an additional condition on the correlation of the Langevin sources $L^{(1)}_{k,LLB-II}$ and $L^{(2)}_{k,LLB-II}$: $\langle L^{(1)}_k(0), L^{(2)}_l(t)\rangle = 0$. This condition is not necessary in the model here since all correlation between the Langevin sources disappear in Eq. (9) due to the orthogonality of the longitudinal and damping components. The use of an additive multiplicative stochastic fields leads to an overestimation of the mean magnetization magnitude as compared to the equilibrium magnetization $m_e$. Moreover, for $T > T_c$, this model may show relatively large mean magnetization magnitudes. This overestimation can be understood by considering that a strong additive noise in multidimensional systems with nonlinearities can generate a random shift far from the deterministic attractor, referred to as a "phantom attractor" [25]. Increasing the volume of the single domain particle reduces the intensity of this additive noise thus removing the effect of the phantom attractor. However, for various applications the particles can be small, and one needs to be able to model their behavior. The proposed formulation has a vanishing longitudinal fluctuations for $T < T_c$, which is similar to the self-consistent Bloch (SCB) formulation in Refs. [10,16]. As compared to SCB formulation, however, the proposed formulation does have longitudinal fluctuations.

## NUMERICAL RESULTS

We implemented the system of equations in a numerical code using the Heun's time integrator. The choice of the scheme is dictated by the Stratonovich interpretation of our model and by the simplicity of implementation [26]. The numerical scheme is identical to the one used for the deterministic equation, where the deterministic part of our equation converges with order 2 and the stochastic part converge with order 1. To study the behavior of our model, we compare the proposed model, referred to as LLB-III, to the LLB model of Garanin (LLB-I) [11] and Evans (LLB-II) [12]. We note that Refs. [11] and [12] use an approximation to the deterministic part of the LLB formulation in terms of the Brillouin function expansion and susceptibilities. For the considered simulations, this approximation leads to almost the same results as those obtained based on the original deterministic part, as defined in Eqs. (5) and (10). For proper comparison in the following results, we used $A_i$ in Eq. (10) in the deterministic part of all the LLB formulations.

To exemplify the model outcomes, we study the magnetization distribution around an equilibrium state for an isotropic and anisotropic single-domain FePt particles. The considered particles has a characteristic length $L = 5\,\text{nm}$ ( $V = L^3 = 125\,\text{nm}^3$ ), Curie Temperature $T_c = 700\,\text{K}$ ( $J_0 = 3k_bT_c$ ), saturation magnetization $M_S^0 = 500\,\text{emu/cm}^3$, magnetic moment of $\mu_0 = 5\mu_B$. Since we are interested in the equilibrium, we use the atomistic damping coefficient of its the critical value $\lambda = 1$, as also chosen by Evans [12]. For the integration scheme we used a time-step of $\Delta t = 1\,\text{fs}$. The magnetization of the system is initially set equal to the equilibrium magnetization obtained from the atomistic model for an ideal SC lattice material [27] and the system is equilibrated for 1ns. The magnetization distribution is obtained from the equilibrated system by sampling the distribution over 10ns.

We first consider an isotropic case (i.e. $\eta'_{x,y} = 0$). Figure 1 shows a uniform distribution of the magnetization with respect to the polar angle $\theta$. This is expected since no energy barrier is present in the direction perpendicular to the magnetization. There is an agreement between the results from all the models.

Figure 2 shows the mean magnetization magnitude as a function of the temperature for particles of different sizes for the three LLB models. The LLB-I model underestimates the mean magnetization magnitude at high temperatures, which is explained by the longitudinal fluctuations. The LLB-II model overestimates the mean magnetization magnitude, which is explained by the effects of the additive noise as discussed after Eq. (24). For $T > T_c$, this overestimation leads to large mean magnetization magnitude values (around 10% of the saturation value), which are significantly higher than the values obtained via the LLB-I and LLB-II models. Increasing the diameter of the particle, the displacement from the equilibrium magnetization disappears and it is negligible for particles of size of $L=20\,\text{nm}$. Since in the LLB-I and LLB-II models the diffusion coefficient is proportional to the inverse of the volume, it means that the equilibrium magnetization is recovered when the diffusion coefficient is reduced. For the LLB-II model, the phantom attractor also disappears if we reduce the intensity of the fluctuations as expected from the theory.

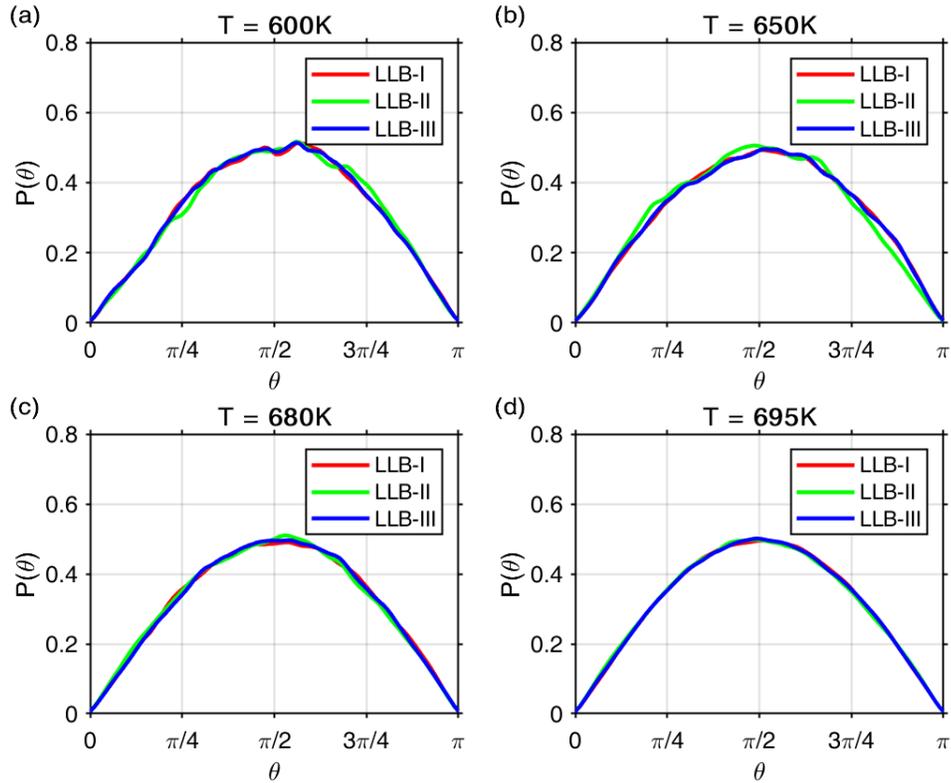

FIGURE 1: DISTRIBUTION OF THE MAGNETIZATION POLAR ANGLE $\theta$ FOR AN ISOTROPIC MAGNETIC PARTICLE ($L=5\,\text{nm}$) FOR THE LLB-I (RED LINES), LLB-II (GREEN LINES), AND LLB-III (BLUE LINES) MODELS. THE RESULTS ARE GIVEN FOR (A) T= 600K, (B) T=650K, (C) T=680K, AND (D) T=695K.

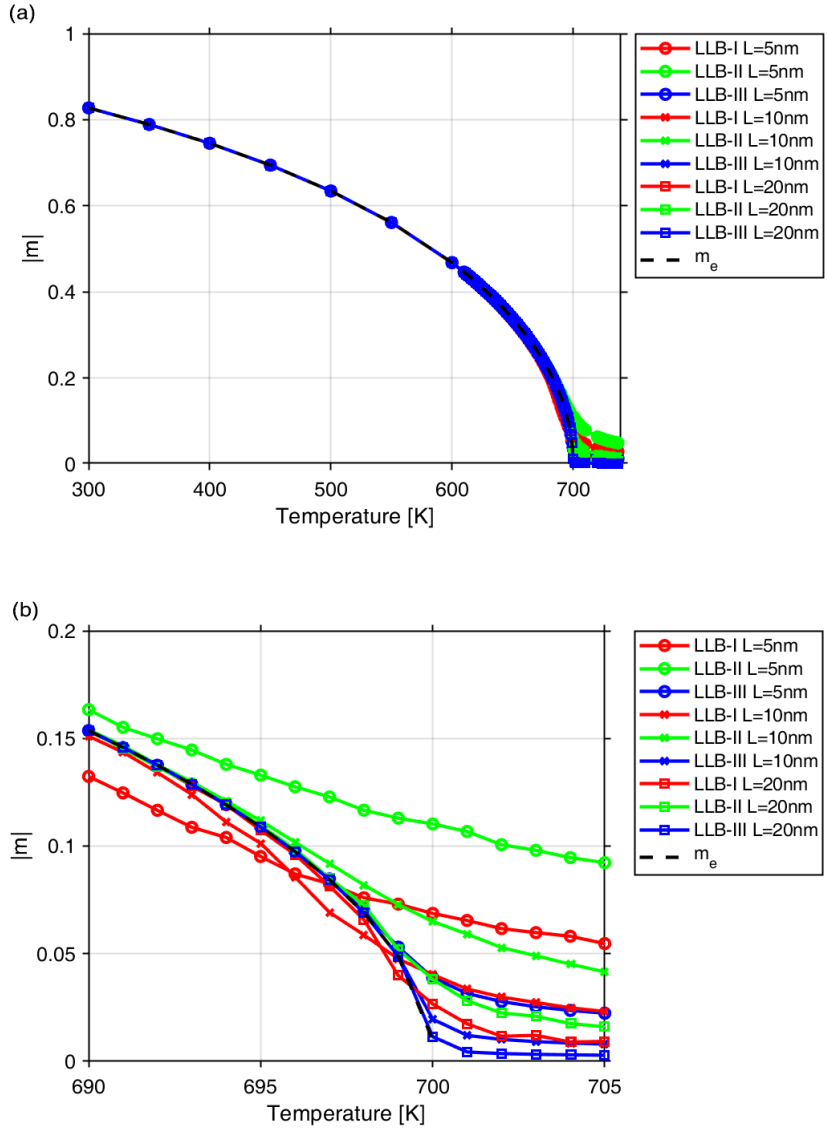

(b)

**FIGURE 2: (A) MAGNETIZATION LENGTH VERSUS TEMPERATURE FOR ISOTROPIC PARTICLES OF DIFFERENT SIZES:** $L = 5\,\text{nm}$ **(CIRCLES),** $L = 10\,\text{nm}$ **(CROSSES), AND** $L = 20\,\text{nm}$ **(SQUARES) FOR THREE MODELS: LLB-I (RED LINE), THE LLB-II (BLUE LINE), AND THE LLB-III (BLUE LINE). THE BLACK DASHED LINE REPRESENTS THE INPUT EQUILIBRIUM MAGNETIZATION** $m_e$ **. (B) INSET FOR A TEMPERATURE RANGE NEAR** $T_c$ **.**

Figures 3 shows the magnetization magnitude probability density for $T > T_c$ for different particle sizes and the three LLB models. We can see that the distribution width obtained using LLB-III is narrower as compared to the results obtained using the LLB-I and LLB-II models. The distributions of the LLB-I and LLB-III models are of MB type as in Eq. (15). The mean values are in agreement with Fig. 2.

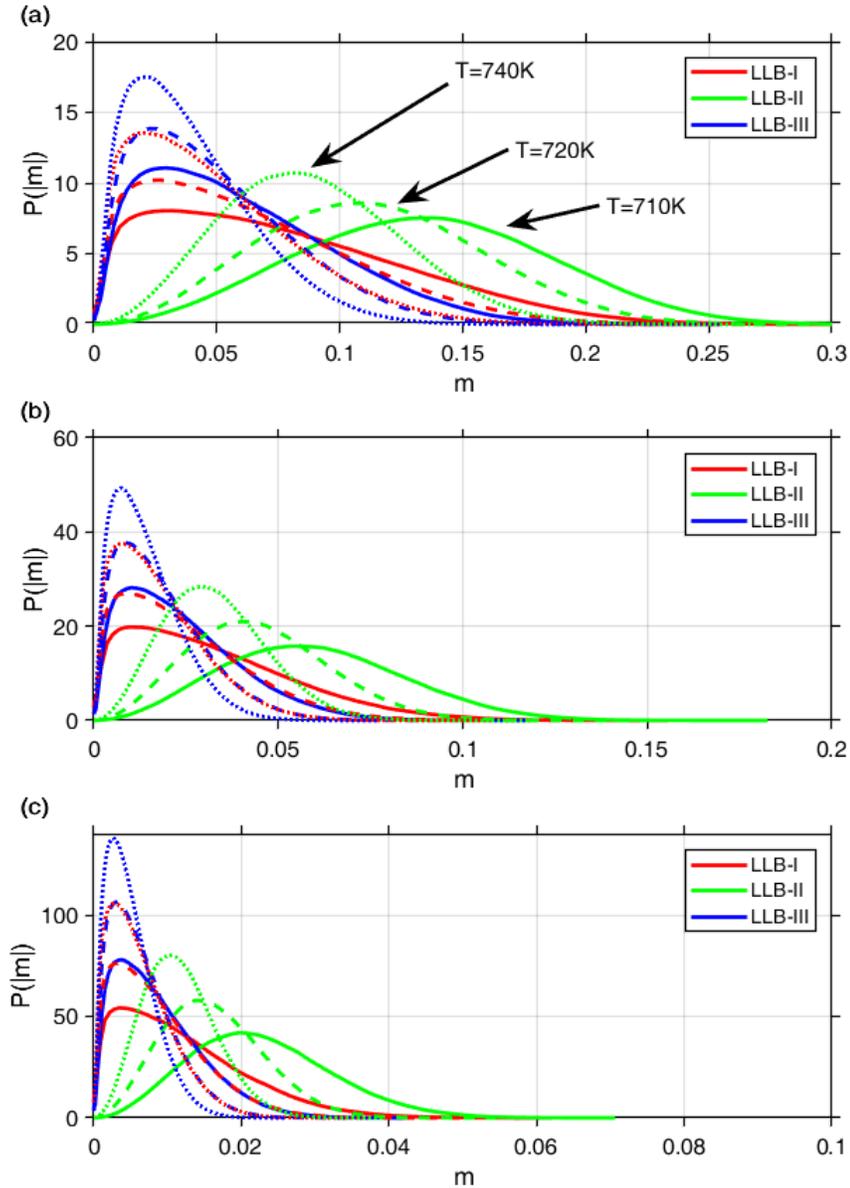

**FIGURE 3: DISTRIBUTION OF THE MAGNETIZATION ABOVE $T_c = 700$K FOR ISOTROPIC PARTICLES OF THREE SIZES: 5 NM IN (A), 10 NM IN (B), AND 20 NM IN (C) USING THE LLB-I (RED LINES), THE LLB-II (GREEN LINES), AND THE LLB-III (BLUE LINES) MODELS. THE RESULTS ARE PROVIDED FOR DIFFERENT TEMPERATURES: $T = 710$K (SOLID LINES), $T = 720$K (DASHED LINES), AND $T = 740$K (DOTTED LINES).**

Next, we consider an anisotropic case with the anisotropy field $H_K = J_0\eta'_{x,y}/\mu_0 = 1.0T$. The angular dependence results for the three models are shown in Fig. 4 The probability distribution along $\theta$ shows two peaks around $\theta = 0$ and $\theta = \pi$ at low temperature (Fig. 4a) due to the presence of the uniaxial anisotropy that gives the magnetization a preferential direction along the anisotropy axis $z$. The two peaks decrease with the temperature (Fig. 4b-c), and the distribution becomes identical to the isotropic case for temperature close to $T_c$ (Fig. 4d). For $T > T_c$, the mean magnetization and the probability density of the magnetization length are qualitatively the same as in Figs. 2 and 3, and therefore, they are not shown.

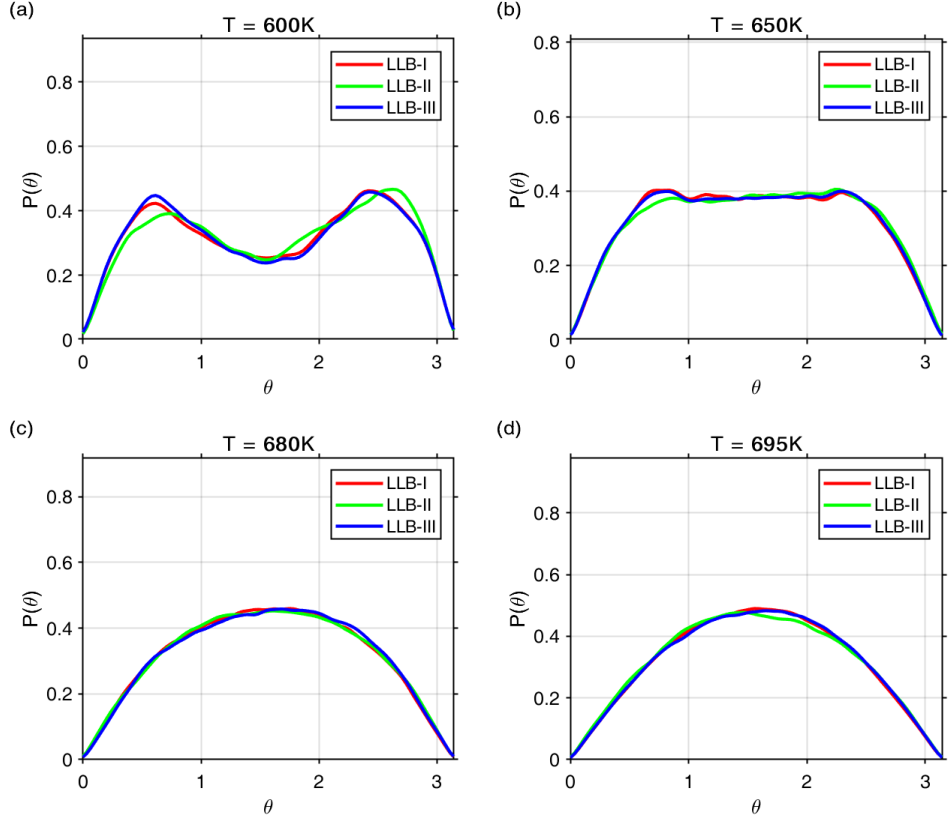

**FIGURE 4: DISTRIBUTION OF THE MAGNETIZATION POLAR ANGLE $\theta$ FOR A MAGNETIC PARTICLE ($L = 5nm$) WITH AN UNIAXIAL ANISOTROPY ALONG Z ($H_k = 1.0T$) FOR THE LLB-I (RED LINES), LLB-II (GREEN LINES), AND LLB-III (BLUE LINES) MODELS. THE RESULTS ARE GIVEN FOR (A) T= 600K, (B) T=650K, (C) T=680K, AND (D) T=695K.**

## SUMMARY

We introduced a formulation for thermal noise in the stochastic form of the LLB equation. The model was tested by considering the equilibrium simulation for small magnetic particles and the results were compared with existing formulations.

The introduced noise formulation is similar to the one proposed by Garanin [11]. The two approaches share common assumptions and obtain the noise using the Fokker-Plank equation. The formulations are different in using different definitions of the free energy of the system. The free energy used here is the same as was used recently in deriving the diffusion coefficients for the SCB equation [10,16]. Using the Fokker-Plank equation, the obtained noise diffusion coefficients lead to a vanishing noise component in the direction of the magnetization for $T<T_c$ and a gradually increasing with $T$ diffusion coefficient for $T>T_c$. An important property of the presented model is that it recovers the expected mean value of the magnetization at the equilibrium for small and large magnetic particles. The distribution of the magnetization length is a MB like distribution. Such a distribution is consistent with the results in other related physics types [19].

The preservation of the equilibrium magnetization makes the presented model appealing for HAMR [22] and multiscale atomistic-micromagnetic modeling [8]. The multiscale modeling would use the Density Functional Theory to provide input parameters for atomistic modeling and the atomistic modeling would provide parameters for the LLB model, which would be used at the mesoscopic scale. The formulation can be used in existing micromagnetic models with a simple update of the noise terms. Even if a direct measure of the longitudinal noise generated is hard to obtain experimentally, an indirect further validation of the proposed model can be obtained by measurements and simulation of HAMR systems.

# APPENDIX: THE FLUCTUATION-DISSIPATION THEOREM

In this work, we have obtained the diffusion coefficient by solving the FP equation at equilibrium. This approach is valid for the entire range of temperature both in the linear and non-linear regimes. The absence of fluctuations at low temperature we obtained in Eq. (23) may seems counter-intuitive. One would expect that the result we obtained be equivalent to the one obtained using the fluctuation dissipation theorem (FDT).

The FDT is a linear-response theory, valid for small deviation from the equilibrium. We base our derivation on the methods described in Refs. [28,29]. At low temperatures and for small fluctuations of the magnetization $\delta m$ near the equilibrium magnetization (i.e. $\delta m \ll m_0$), we can linearize Eq. (6), converting the multiplicative noise into an additive noise variable. The linearized system of equation can then be written in the absence of thermal fluctuations as:

$$\frac{dx_i}{dt} = \sum_{j=1}^{3N} L_{ij} x_j \ , \qquad (A1)$$

where $i=1,\ldots,3N$ are the degree of freedom of the $N$ particles in the system, $x_i = \delta m_i$ is the deviation of the subsystem $i$ from the equilibrium, and $L_{ij}$ are the components of the linearize matrix $\mathbf{L}$. The general (linearized) Langevin equation of motion is written in the form:

$$\frac{dx_i}{dt} = -\sum_{j=1}^{3N} \gamma_{ij} X_j + f_i \ , \qquad (A2)$$

where $\gamma_{ij}$ are the kinetic coefficients, $f_i$ represent random forces responsible for the spontaneous fluctuation, and $X_j$ are thermodynamically conjugate variables related to the entropy $s$ of the magnetic system by:

$$X_j = -\frac{\partial S}{\partial x_j} \ . \tag{A3}$$

For a closed system in an external medium, Eq. (A3) can be written as:

$$X_j = \beta \frac{\partial F}{\partial x_j} \ , \tag{A4}$$

where $F$ is the free energy of the system, defined in Eq.(7) that can be expressed in terms of the $x_i$ as:

$$F = F_0 + \frac{1}{2} \sum_{ij} A_{ij} x_i x_j \ , \tag{A5}$$

where $A_{ij}$ are the components of the symmetric energy matrix and $F_0$ is a constant. Defining the field variations due to the small fluctuation of the magnetization as $h_i = h_i(x_1, x_2, \cdots, x_{3N}) = \sum_{ij} B_{ij} x_j$, we can rewrite Eq. (A4) as:

$$X_j = -\beta M_s V h_j = -\beta M_s V \sum_{ij}^{3N} B_{ij} x_j \ . \tag{A6}$$

The statistical properties of the random forces $f_i$ in Eq. (A2) can be obtained using the Onsager principle:

$$\langle f_i(t) \rangle = 0 \quad \langle f_i(0) f_j(t) \rangle = (\gamma_{ij} + \gamma_{ji}) \delta(t) \ . \tag{A7}$$

The kinetic coefficient can be obtained by solving:

$$\frac{dx_i}{dt} = \sum_{j=1}^{3N} L_{ij} x_j = \beta M_s V \sum_{j=1}^{3N} \gamma_{ij} h_j \ . \tag{A8}$$

Thus, the kinetic coefficient can be obtained as:

$$\gamma_{ij}^{\mu\nu} = \frac{L_{ij}^{\mu\nu}}{\beta M_s V B_{ij}^{\mu\nu}} \ , \tag{A9}$$

where $i, j = 1, \ldots, N$ represent the $N$ macrospin in our system, and $\mu\nu = x, y, z$ are the cartesian coordinates.

We consider an initial magnetization $\mathbf{m}_i = \{0, 0, m_0\}$ with $i = 1, \ldots, N$, and we introduce small fluctuations of the magnetization (i.e. $\delta m_i^x, \delta m_i^y, \delta m_i^y \ll m_0$). The components of the field matrix $B_{ij}$ can be obtained from Eq. (2) and Eq. (7) as:

$$B_{ij}^{xx} = \left[ \frac{A_{ex}}{h^2} (\delta_{ij-1} - 2\delta_{ij} + \delta_{ij}) - H_k \delta_{ij} \right] , \tag{A10}$$

$$B_{ij}^{yy} = \left[ \frac{A_{ex}}{h^2} (\delta_{ij-1} - 2\delta_{ij} + \delta_{ij}) - H_k \delta_{ij} \right] . \tag{A11}$$

$$B_{ij}^{zz} = \left[ \frac{A_{ex}}{h^2} \left( \delta_{ij-1} - 2\delta_{ij} + \delta_{ij} \right) + \frac{J_0}{\mu_0} \delta_{ij} \right] , \tag{A12}$$

and $B_{ij}^{\mu\nu} = 0$ for $\mu \neq \nu$.

The linearized equation of the magnetization can be expressed in the form:

$$L_{ij}^{xx} = \frac{\gamma \tilde{\alpha}_\perp}{m_0^2} \left[ \frac{A_{ex}}{h^2} \left( \delta_{ij-1} - 2\delta_{ij} + \delta_{ij} \right) - H_k \delta_{ij} \right] , \tag{A13}$$

$$L_{ij}^{yy} = \frac{\gamma \tilde{\alpha}_\perp}{m_0^2} \left[ \frac{A_{ex}}{h^2} \left( \delta_{ij-1} - 2\delta_{ij} + \delta_{ij} \right) - H_k \delta_{ij} \right] , \tag{A14}$$

$$L_{ij}^{zz} = \frac{\tilde{\alpha}_\parallel}{m_0^2 \beta \mu_0 B'(\xi_0)} \left[ B_s(\xi_0) + m_0 \frac{\delta \xi_0}{\delta m} B'(\xi_0) - 2 m_0 \right] , \tag{A15}$$

$$L_{ij}^{xy} = -L_{ij}^{yx} = -\gamma \left[ \frac{A_{ex}}{h^2} \left( \delta_{ij-1} - 2\delta_{ij} + \delta_{ij} \right) - H_k \delta_{ij} \right] , \tag{A16}$$

$$L_{ij}^{xz} = L_{ij}^{yz} = L_{ij}^{zx} = L_{ij}^{zy} = 0 . \tag{A17}$$

In equation (A15), $\delta \xi_0 = \mu_0 \beta h$ is the fluctuation of the reduced magnetization due to a small change in the magnetization from the equilibrium value. Using the relationship given in Eq.(3), it is easy to show that:

$$\frac{\delta m}{\delta \xi_0} = \frac{\delta B_s(\xi_0)}{\delta \xi_0} = B'_s(\xi_0) . \tag{A18}$$

Using Eq. (A18) and Eq.(3) in Eq. (A15), we can show that $L_{ij}^{zz} = 0$. Hence, we can see that the longitudinal fluctuation does not exist in our system in the linearized case. This result is consistent we the result we obtained using the FP equation below the Curie Temperature.